\magnification=1220
\baselineskip=15pt
\def\medskip{\vskip .2in}

\def\ds{\displaystyle}
\pageno=1
\footline{\hss\folio\hss}
\vskip2.5pc
\centerline{\bf  Constant of Motion for One-Dimensional Non Autonomous Linear}
\centerline{\bf  Systems and Harmonic Oscillator} 
\vskip4pc
\centerline{G. L\'opez}
\vskip1pc
\centerline{Departamento de F\'isica de la Universidad de Guadalajara}
\centerline{Apartado Postal 4-137}
\centerline{44410~Guadalajara, Jalisco, M\'exico}
\vskip5pc
\centerline{August 1999}
\vskip1pc
\centerline{PACS~~03.20.+i~~~03.65.Ca}

\vskip4pc
\centerline{ABSTRACT}
\vskip2pc
 For a one-dimensional motion, a constant of motion for non autonomous an
linear  system (position and velocity) is given from the constant of motion
associated to its autonomous system. This approach is used in the study of the
harmonic oscillator with an additional time depending force.
  \vfil\eject
\line{\bf I. Introduction\hfil}
\vskip1pc
The importance of the constant of motion of an autonomous system (the forces
are time independent) and its relation with the Lagrangian and the Hamiltonian
is well known (L\'opez, 1996). For autonomous systems, the constant of motion ,
Lagrangian and Hamiltonian do not need to depend explicitly on time (Goldstein,
1950). If one looks for these quantities to be time explicitly depending, one
could get ambiguities in the formulation (L\'opez, 1998). On the other hand,
constant of motion for nonautonomous systems (the forces depend explicitly on
time) is very little known, and the usual approach to study these systems is
looking for a Hamiltonian without worry about whether or not it is a constant of
motion (Zel'dovich, 1967; Dekker,1981,and references therein). A constant of
motion  of nonautonomous system must depend explicitly on time and has its own
importance by its own, independently on a possible relation with the Lagrangian
and Hamiltonian of the system (L\'opez and Hern\'andez, 1987). In this paper, a
constant of motion for a nonautonomous system which has a linear dependence on
coordinate and velocity is deduced. This approach is applied to the harmonic
oscillator with dissipation and with a time depending force.
\vskip3pc
\line{\bf II. Constant of Motion\hfil}
\vskip1pc
Consider the following nonautonomous dynamical system
$${\ds dx\over\ds dt}=v\eqno(1a)$$
and
$${\ds dv\over\ds dt}=ax+bv+f(t)\ ,\eqno(1b)$$
where $a$, $b$ are constants and $f(t)$ is an arbitrary function. For a
constant of motion associated to this system, one understands a function
$K=K(x,v,t)$ such that its total derivation with respect the time is zero
($dK/dt=0$), that is, $K$ must be solution of the following partial
differential equation
$$v~{\ds\partial K\over\ds\partial x}+\biggl(ax+bv+f(t)~\biggr)~{\ds\partial K
\over\ds\partial v}+{\ds\partial K\over\ds\partial t}=0\ .\eqno(2)$$
Assume that $K_o=K_o(x,v)$ is the constant of motion associated to (1) but for
$f=0$ (associated autonomous system), that is, $K_o$ is solution of th
equation
$$v~{\ds\partial K_o\over\ds\partial x}+(ax+bv)~{\ds\partial K_o\over\ds
\partial v}=0\ .\eqno(3)$$
Thus, it follows that
$$K(x,v,t)=K_o\biggl(x-\alpha(t),v-{d\alpha\over dt}\biggr)\eqno(4)$$
is solution of Eq. (2), where $\alpha(t)$ is the particular solution of the
equation
$${\ds d^2\alpha\over\ds dt^2}=a\alpha+b~{\ds d\alpha\over\ds dt}+f(t)\
,\eqno(5)$$
To see this, defining the variables $\xi_1$ and $\xi_2$ as
$$\xi_1=x-\alpha(t),\quad\quad\xi_2=v-d\alpha/dt\ ,\eqno(6)$$
one gets the following derivations
$$v~{\ds\partial K\over\ds\partial x}=v~{\ds\partial K_o\over\ds\partial\xi_1},
\eqno(7a)$$
$$(ax+bv+f)~{\ds\partial K\over\ds\partial v}=(ax+bv+f)~{\ds\partial K_o\over\ds
\partial\xi_2}\ ,\eqno(7b)$$
and
$${\ds\partial K\over\ds\partial t}=-{\ds d\alpha\over\ds dt}~{\ds\partial K_o
\over\ds\partial\xi_1}-{\ds d^2\alpha\over\ds dt^2}~{\ds\partial
K_o\over\ds\partial \xi_2}\ .\eqno(7c)$$
Substituting Eqs. (7a), (7b), and (7c) in Eq. (2), rearranging terms, and using
Eqs. (5) and (6), one has
$${\ds dK\over\ds dt}=\xi_2~{\ds\partial K_o\over\ds\partial\xi_1}+
\bigl[a\xi_1+b\xi_2\bigr]~{\ds\partial K_o\over\ds\partial\xi_2}=0\eqno(8)$$
since by definition (3), $K_o$ satisfies this type of equation. Therefore,
(4) is solution  of  Eq. (2) and is a constant of motion of the
nonautonomous system (1).
\vskip3pc
\leftline{\bf III. Harmonic Oscillator with Periodic Force}
\vskip1pc
This dynamical system is defined by the equations
$${\ds dx\over\ds dt}=v\eqno(9a)$$
and
$${\ds dv\over\ds dt}=-\omega^2+{A\over m}\sin(\Omega t)\ ,\eqno(9b)$$
where $\omega$ is the natural frequency of oscillations, $m$ is the mass of the
particle, $\Omega$ and $A$ are the frequency and amplitude of the external
periodic force. According to Eqs. (4) and (5), the function $\alpha$ is the
particular solution of Eq. (9) which is given by
$$\alpha(t)=-{\ds A\sin(\Omega t)\over\ds m(\Omega^2-\omega^2)}\ .\eqno(10)$$
Therefore, a constant of motion of system (9) can be written for
$\Omega\not=\omega$ (out of resonance) as
$$\eqalign{K(x,v,t)&=
{1\over 2}mv^2+{1\over 2}m\omega^2x^2\cr
&+{\ds A\over\Omega^2-\omega^2}\biggl(\Omega v\cos(\Omega t)+
\omega^2 x\sin(\Omega t)\biggr)\cr
&-{\ds A^2\over\ds 2m(\Omega^2-\omega^2)}\sin^2(\Omega t)\ ,\cr}\eqno(11)$$
where a term of the form $A^2\Omega^2/2m(\Omega^2-\omega^2)$ has been ignored since
it represents a constant term.

For the resonant case ($\Omega=\omega$), one gets
$$\alpha(t)={\ds A\over\ds 4m\omega^2}\sin(\omega t)-{\ds A\over 2m\omega}~
t\cos(\omega t)\ ,\eqno(12)$$
and the constant of motion can be written as
$$\eqalign{
K(x,v,t)&={1\over 2}mv^2+{1\over 2}m\omega^2x^2\cr
&+{\ds A\over\ds 4\omega}\biggl[(v+x\omega^2t)\cos(\omega t)-
(x\omega+2v\omega t)\sin(\omega t)\biggr]\cr
&+{\ds A^2t\over\ds 8m\omega}[\omega t-\sin(2\omega t)~]\ ,\cr}\eqno(13)$$
where a term of the form $A^2/32m\omega^2$ has been ignored. It is not difficult
to see that expression (12) satisfies indeed the following equation
$$v~{\ds\partial K\over\ds\partial x}+\biggl[-\omega^2x+{\ds A\over m}\sin(\Omega
t) \biggr]{\ds\partial K\over\ds\partial v}+{\ds\partial K\over\ds\partial t}
=0\ .\eqno(14)$$
\vfil\eject
\leftline{\bf IV. Dissipative Harmonic Oscillator with Periodic Force}
\vskip1pc
This dynamical system is defined by the equations
$${\ds dx\over\ds dt}=v\eqno(15a)$$
and
$${\ds dv\over\ds dt}=-\omega^2x-{\ds\lambda\over\ds m}~v+{\ds A\over\ds
m}~\sin(\Omega t)\ ,\eqno(15b)$$
where $\lambda$ is the parameter which characterizes the dissipation. The particular solution of
this system is given by
$$\alpha(t)={\ds A/m\over\left({\ds\lambda\Omega\over\ds m}\right)^2+
(\omega^2-\Omega^2)^2}~\left[(\omega^2-\Omega^2)\sin(\Omega t)-{\ds\lambda\Omega
\over\ds m}~\cos(\Omega t)~\right]\ .\eqno(16)$$
System (15) is also of the form (1), and the constant of motion for $A=0$ (associated
autonomous system) was given somewhere else (L\'opez, 1996). Thus, according to this reference
and Eq. (4), a constant of motion of the system (15) can be given by
$$\eqalign{
K(x,v,t)&={\ds m\over\ds 2}\left[(v-\beta(t))^2+{\ds\lambda\over\ds
m}(x-\alpha(t))(v-\beta(t))+\omega^2(x-\alpha(t))^2\right]\cr
&\times\exp\left(-{\lambda\over m}G\biggl({v-\beta(t)\over x-\alpha(t)},w,\lambda
\biggr)\right)\ ,\cr}\eqno(18a)$$
where $G=G(\xi,w,\lambda)$ is the function defined as

$$G=\cases{{\ds 1\over\ds 2\sqrt{(\lambda/2m)^2-\omega^2}}
\log {\ds\lambda/2m+\xi-\sqrt{(\lambda/2m)^2-\omega^2}\over\ds
\lambda/2m+\xi+\sqrt{(\lambda/2m)^2-\omega^2} },
&if~~$\omega^2<(\lambda/2m)^2$\cr\cr
{\ds 1\over\ds \lambda/2m+\xi},&if~~$\omega^2=(\lambda/2m)^2$\cr\cr
{\ds 1\over\ds\sqrt{\omega^2-(\lambda/2m)^2}}\arctan{\lambda/2m+\xi\over\ds
\sqrt{\omega^2-(\lambda/2m)^2}},
&if~~$\omega^2>(\lambda/2m)^2$\cr}\eqno(18c)$$
 and the function $\beta(t)$ is defined as
$$\beta(t)={\ds A\Omega/m\over\left({\ds\lambda\Omega\over\ds m}\right)^2+
(\omega^2-\Omega^2)^2}~\left[(\omega^2-\Omega^2)\cos(\Omega t)+{\ds\lambda\Omega
\over\ds m}~\sin(\Omega t)~\right]\ .\eqno(18c)$$
For the particular case of very weak dissipation ($\lambda/2m\ll\omega$), one
gets the expression
$$\eqalign{
K(x,v,t)&={\ds m\over 2}\biggl[(v-\beta(t))^2+\omega^2(x-\alpha(t))^2\biggr]
+{\ds\lambda\over 2}(x-\alpha(t))(v-\beta(t))\cr
&-{\ds\lambda\over 2\omega}\biggl[(v-\beta(t))^2+\omega^2(x-\alpha(t))^2\biggr]
\arctan{v-\beta(t)\over\omega(x-\alpha(t))}\ .\cr}\eqno(19)$$
\vskip4pc
\leftline{\bf V. CONCLUSION}
\vskip1pc
For a particle moving in one dimension where forces are linear in position and
velocity, to know a constant of motion for this system (autonomous) allows one
to get a constant of motion when a time-depending force is added. This approach
was applied to the harmonic oscillator with external force and dissipation.
\vfil\eject
\leftline{\bf References}
\vskip2pc
\obeylines{
Goldstein, H. (1980) Classical Mechanics, Addison-Wesley,Reading, Massachusetts.
Dekker, H. (1981). {\it Physics Reports}.{\bf 52}. 263.
L\'opez, G.,(1996){\it Annals of Physics}. {\bf 251}(2).372.
L\'opez, G.,(1998){\it International Journal of Theoretical Physics}.{\bf 37}(5).1617.
L\'opez, G. and Hern\'andez, J. I.(1987). {\it Annals of Physics}. {\bf 193}.1.
Zel`dovich, Ya. B. (1967).{\it Soviet Physics JETP}. {\bf 24}(5). 1006.
}
\end